\documentclass[prl,superscriptaddress,twocolumn,nopreprintnumbers,
floatfix,nofootinbib]{revtex4}

\usepackage{graphicx,url,amssymb,amsmath,longtable,rotating,color,units,
wasysym,subfigure,epsfig,multirow,ulem}

\usepackage[plainpages=false, colorlinks=true, anchorcolor=blue, linkcolor=blue, citecolor=blue, bookmarks=false]{hyperref}

\newcommand{\LL}{{\mathcal L}}

\begin{document}

\title{Detecting gravitational-wave memory with LIGO: implications of GW150914}
\author{Paul D. Lasky}
\email{paul.lasky@monash.edu}

\author{Eric Thrane}
\author{Yuri Levin}

\affiliation{Monash Centre for Astrophysics, School of Physics and Astronomy, Monash University, VIC 3800, Australia}

\author{Jonathan Blackman}
\affiliation{TAPIR, Walter Burke Institute for Theoretical Physics, California Institute of Technology, Pasadena, California 91125, USA}

\author{Yanbei Chen}
\affiliation{Theoretical Astrophysics, California Institute of Technology, Pasadena, CA 91125, USA}


\begin{abstract}
It may soon be possible for Advanced LIGO to detect hundreds of binary black hole mergers per year.
We show how the accumulation of many such measurements will allow for the detection of gravitational-wave memory: a permanent displacement of spacetime that comes from strong-field, general relativistic effects.  We estimate that Advanced LIGO operating at design sensitivity may be able to make a signal-to-noise ratio 3(5) detection of memory with $\sim 35\,(90)$ events with masses and distance similar to GW150914.  
We highlight the importance of incorporating higher-order gravitational-wave modes for parameter estimation of binary black hole mergers, and describe how our methods can also be used to detect higher-order modes themselves before Advanced LIGO reaches design sensitivity.
\end{abstract}

\maketitle

The landmark detection of gravitational waves (GWs)
indicates binary black hole mergers will soon be observed regularly with Advanced LIGO (aLIGO) ~\cite{abbott16_detection}.
This opens the door to new areas of study not possible with light given the electromagnetically quiet nature of black hole mergers.  We turn our attention to Christodoulou GW memory~\cite{zeldovich74,braginsky87,christodoulou91,thorne92,favata09a,bieri14}, a purely strong-field gravitational effect.  We show memory can be probed in the near future using an ensemble of observations of binary black hole systems.

Memory induces a monotonically increasing GW strain, a permanent change in the relative distance between two freely falling test masses~\cite{zeldovich74, braginsky87}.  Pulsar timing arrays regularly search for memory \cite[][]{pshirkov10,vanhaasteren10,seto09,wang15,arzoumanian15_memory}.  However, the memory component is a small fraction of the total strain for a merger, making it improbable that aLIGO will detect memory from an individual coalescence signal~\cite[][]{favata09a}.  The optimal, matched filter memory signal-to-noise ratio for an event like GW150914 with two aLIGO detectors at design sensitivity averaged over astrophysical parameters is $\mbox{S/N}=0.42$.  Although this is too small to be detected, it is non-negligible.  However, an {\it ensemble} of events can be used to detect memory.  Moreover, GW150914 implies mergers are relatively abundant in the Universe, and therefore GW observations are likely to become commonplace in the near future~\cite{abbott16_rates}.

The production of memory waveforms from numerical relativity simulations is in its infancy \cite{pollney11}.  Analytic waveforms that use a variety of approximations match numerical simulations qualitatively.  Both techniques yield a step-function-like waveform with comparable rise time and amplitude \cite[][]{favata09a, favata10, pollney11}.  We use the {\it minimal-waveform model} \cite{favata09b}.  This phenomenological model has been matched to calculations using both effective-one-body approach~\cite{favata09b} and with pure numerical relativity simulations~\cite{pollney11}.  We use this model because it is simple, and provides qualitative agreement with waveforms derived by other means.  

For non-spinning, quasi-circular binaries, the memory waveform is linearly polarized, and is described by the following parameters: masses ($m_1$, $m_2$), distance to the source $d$, inclination angle $\theta$, and polarization angle $\psi$.
The assumption of no spin is consistent with GW150914~\cite{abbott16_detection,abbott16_PE}, although the formalism can be straightforwardly extended. However, for strong precession where the orbital plane precesses, the memory is not linearly polarized, complicating the analysis.  Memory waveforms for equal-mass binaries with aligned spins can be found in~\cite{pollney11}.  

The detection of the coalesence's oscillatory component favours face-on binaries, where the binary's orbital angular momentum vector is aligned with the observer's line of sight, $\theta=0$.  However, the memory component of the strain scales as $h^{(\mbox{mem})}\propto\sin^2\theta(17+\cos^2\theta)$ \cite{favata09b}, implying face-on binaries have zero memory, and edge-on binaries have maximal memory.  Our prescription for detecting memory requires each binary to be detected through its oscillatory component.  Parameter estimation for GW150914 yields $\theta\approx140^\circ$, but is consistent with being a face-on/off system~\cite{abbott16_PE}.  Comparing the memory amplitude for $\theta=140^\circ$ and $\theta=90^\circ$, we find that the signal from a $\theta=140^\circ$ binary is $43\%$ the maximum possible.  The angle $\theta$ does {\it not} affect the memory in any way other than its amplitude.

The polarization modulates the amplitude of the memory signal and the sign.  Our strategy for detecting memory relies on the coherent summation of an ensemble of sub-threshold signals.  We require knowledge of the {\it sign} of the memory for individual detections, otherwise the memory adds incoherently and cancels\footnote{Strictly speaking, it is still possible to measure a signal from an incoherent sum, but this grows much slower.}.  However, it turns out that the sign of the memory cannot be determined using the $h_{\ell m}=h_{22}$ mode of the oscillatory signal that is normally used in parameter estimation~\cite[][]{aasi14d,abbott16_PE}.  Fortunately, the sign can be determined using higher-order modes.

Using the $h_{22}$ mode, the oscillatory waveform is invariant under a simultaneous rotation of the polarisation angle, $\psi\rightarrow\psi+\pi/2$, and a shift in the phase at coalescence, $\phi_c\rightarrow\phi_c+\pi/2$.  That is $h_{22}(\psi,\,\phi_c)=h_{22}(\psi+\pi/2,\,\phi_c+\pi/2)$.  However, memory acquires a minus sign under the same transformation: $h_{\rm mem}(\psi,\,\phi_c)=-h_{\rm mem}(\psi+\pi/2,\,\phi_c+\pi/2)$.  This degeneracy between $\psi$ and $\phi_c$ implies we cannot know the sign of the memory using only the $h_{22}$ component.

\begin{figure}
\begin{center}
	\includegraphics[width=0.9\columnwidth]{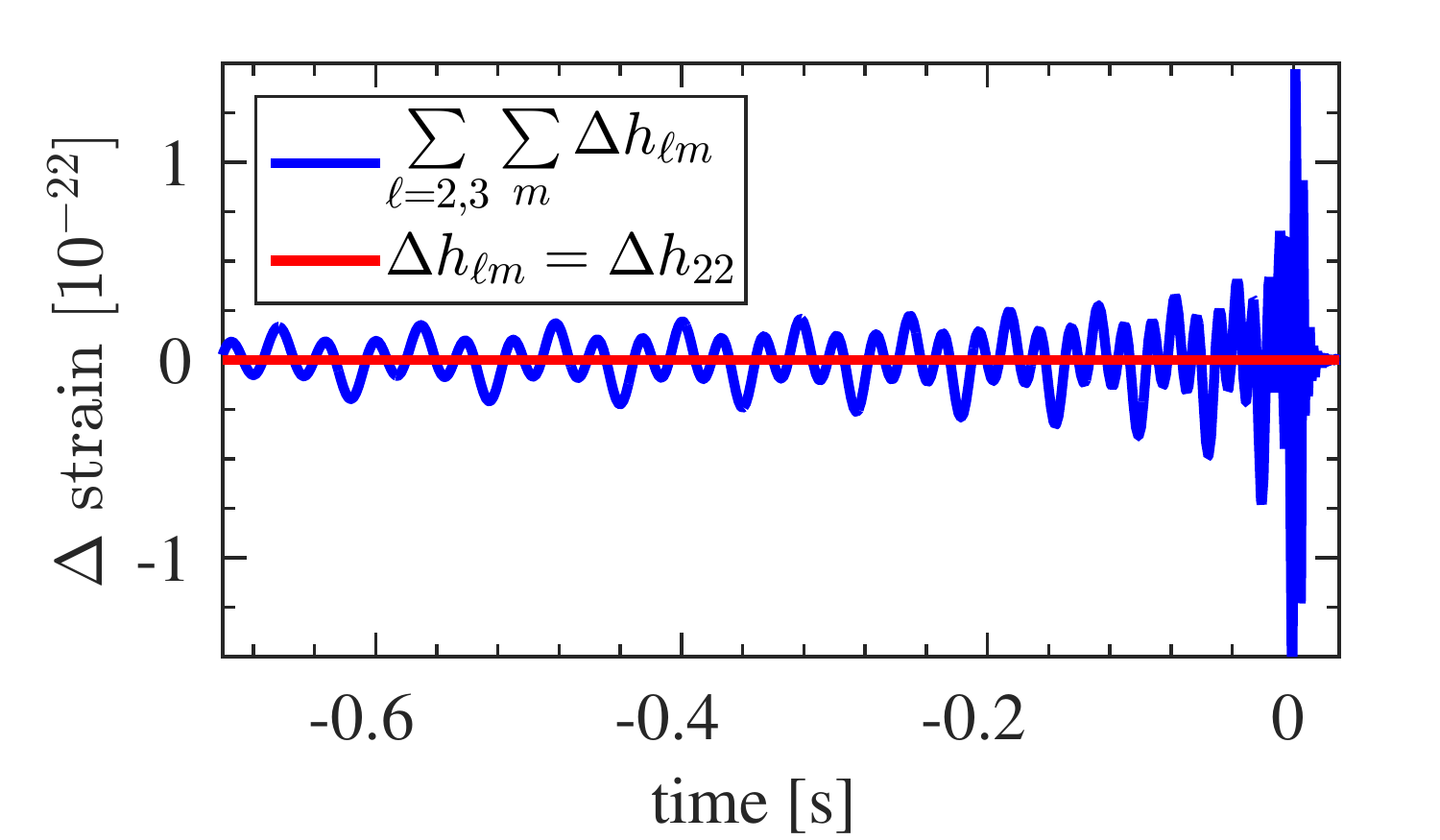}
	\caption{\label{higher_modes} Gravitational-wave time series of the higher-order modes for an edge-on binary with parameters consistent with GW150914~\cite{abbott16_detection,abbott16_PE}, where the vertical axis $\Delta h_{\ell m}$ is defined in Eqn.~(\ref{delta_strain}).  The red curve shows the $\Delta h_{22}$ mode, which is identically zero, implying the GW polarization angle $\psi$ and phase at coalescence $\phi_c$ are degenerate variables.  The blue trace shows $\sum_\ell\sum_m\Delta h_{\ell m}$ for $\ell=2,3$ and all corresponding values of $\left|m\right|>0$.  The fact that $\Delta h_{\ell m}\neq0$ implies that higher-order modes can be used to break the $\psi$ degeneracy and thus determine the sign of the memory.}
\end{center}
\end{figure}

Higher-order $h_{\ell m}$'s can be used to break the degeneracy between $\psi$ and $\psi+\pi/2$; see Fig.~\ref{higher_modes}.  We calculate waveforms using surrogate models~\cite{field14} that include all modes up to $\ell=3$~\cite{blackman15}.  We define 
\begin{equation}
	\Delta h_{\ell m}\equiv \left[h_{\ell m}(\psi,\,\phi_c) - h_{\ell m}(\psi+\pi/2,\,\phi_c+\pi/2)\right]{}_{-2}Y_{\ell m},\label{delta_strain}
\end{equation}
where ${}_{-2}Y_{\ell m}$ are the spin-weighted spherical harmonics.  One can think of $\Delta h_{\ell m}$ as a degeneracy-breaking parameter; a measurable $\Delta h_{\ell m}(t)$ breaks the degeneracy between $\psi$ and $\psi+\pi/2$, and determines the sign of the memory\footnote{In principle, uncertainty in parameters such as component masses and inclination angle reduces our ability to measure $\Delta h_{\ell m}$.  Using a Monte Carlo simulation, we estimate only an $\approx2\%$ systematic error in $\Delta h_{\ell m}$ on average, suggesting this is a small effect.}.

The red trace in Fig.~\ref{higher_modes} shows $\Delta h_{22}=0$, because the $\ell=m=2$ mode does {\it not} break the degeneracy.  The blue curve shows $\sum_\ell\sum_m\Delta h_{\ell m}$ for $\ell=2,\,3$ and all corresponding $\left|m\right|>0$.  The $\psi$ degeneracy is broken and the sign of the memory determined when $\Delta h_{\ell=(2,3) m}$ (blue curve) is detectable with matched filtering. 

\begin{figure}
\begin{center}
	\includegraphics[width=0.9\columnwidth]{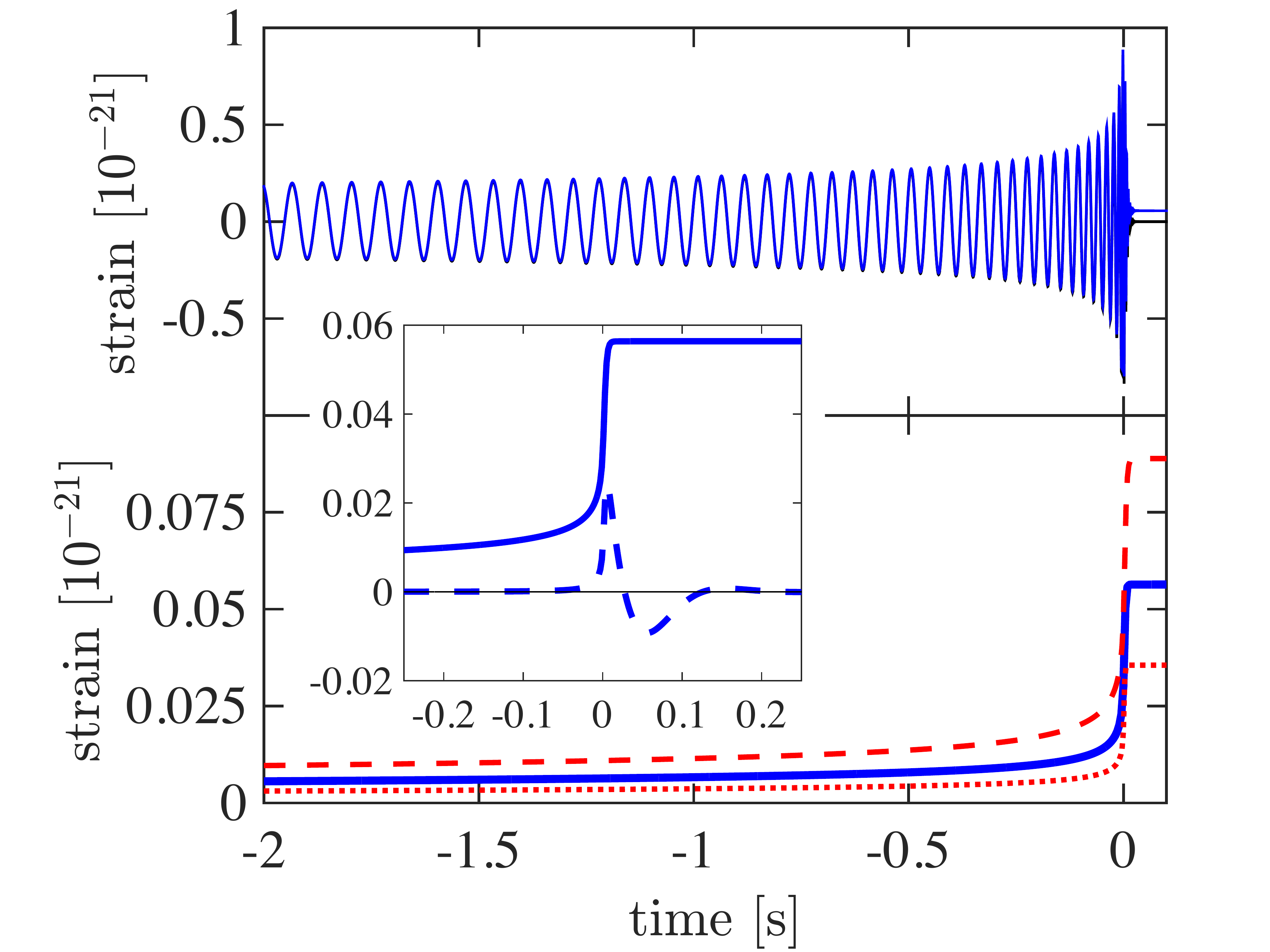}
	\caption{\label{time_series} Gravitational-wave strain time series using parameters consistent with GW150914 \cite{abbott16_detection,abbott16_PE}.  The top panel shows the strain time series with GW memory (blue curve) and without (black).  The bottom panel shows only the memory-induced strain series, where the blue curve uses the maximum likelihood parameters for GW150914 \cite{abbott16_detection, abbott16_PE}.  The red dotted and dashed curves are binaries at the same distance (410 Mpc) and with the same orientation ($\theta=140^\circ$), but equal mass binaries with $m_{1,2}=20\,M_\odot$ and $50\,M_\odot$ respectively (cf. $65\,M_\odot$ for the blue curve).  Inset: the solid blue curve shows a zoomed-in version of the blue curve from the bottom panel, while the dashed curve is after a high-pass filter to show the signal visible in aLIGO.}
\end{center}
\end{figure}

In Fig.~\ref{time_series}, we plot the strain time series for a binary with parameters equivalent to the maximum-likelihood estimates for GW150914: $m_1=36\,M_\odot$, $m_2=29\,M_\odot$,  $d=410\,\mbox{Mpc}$ and $\theta=140^\circ$~\cite{abbott16_detection,abbott16_PE}.  The top panel shows the full signal with and without memory (blue and black curves, respectively).  The bottom panel shows only the memory component.  The memory component is calculated using Eqn.~(9) from Ref.~\cite{favata09b} and the method described therein.  The red dotted and dashed curves are binaries at the same distance, and with the same orientation, but with different masses: $m_{1,2}=20\,M_\odot$ (dotted curve) and $50\,M_\odot$ (dashed curve).  

LIGO is not sensitive to strain below $\approx10$ Hz.  The solid blue curve in the inset to Fig.~\ref{time_series} shows a zoomed-in version of the GW strain corresponding to the blue curve in the bottom panel, while the dashed curve shows the signal after applying a high-pass filter with a 10 Hz cut-off.

The memory amplitude scales linearly with the black-hole masses.  The mass also changes the memory rise time, and hence its spectral shape, but only at frequencies greater than one over the rise time.  LIGO is more sensitive to higher mass binaries providing the characteristic rise time of the memory signal is smaller than $1/f_0$, where $f_0$ is the detector's low-frequency seismic cut-off.

We calculate a matched-filter signal-to-noise ratio, S/N.  For a strain time series $h(t)$ the S/N is
\begin{align}
	\mbox{S/N}=\left<h,\,u\right>/\sqrt{\left<u,\,u\right>},\label{SNR}
\end{align}
where $u(t)$ is the template, and
\begin{align}
	\left<a,\,b\right>\equiv4{\rm Re}\int_0^\infty\frac{\tilde{a}(f)\tilde{b}^\star(f)}{S_h(f)}df,
\end{align}
where $S_h(f)$ is the noise power spectral density.

This S/N calculation assumes the oscillatory component of the coalescence has been successfully removed from the $h(t)$ and that our knowledge of the memory waveform is complete.  We test how this latter assumption affects our ability to measure S/N.  We create memory signals with maximum-likelihood parameters of GW150914, and calculate the S/N expectation value recovered using mismatched templates with $m_{1,2}=46,\,39\,M_\odot$ and $m_{1,2}=26,\,19\,M_\odot$ (note these errors are much larger than the 90\% CL intervals of GW150914).  We find at worst, the S/N expectation value decreases by $\approx21\%$. 

There are good reasons to suppose that the imperfect subtraction of the oscillatory component of the signal does not significantly affect our results.  Firstly, the memory signal is loudest during the merger phase.  It is dominated by the lowest frequency component in the observing band. During the merger, the oscillatory component is about an order of magnitude higher in frequency, suggesting that the residuals from imperfect subtraction will not contribute significantly to the memory signal.  Secondly, as we discuss below, the memory strain adds coherently whereas residual strain from imperfect subtraction adds incoherently.  Therefore, we do not expect uncertainty associated with the binary parameters to significantly affect our results.

For multiple events, we construct an optimal estimator for the weighted-memory sum
\begin{align}
	\hat{h}_{\rm tot} =& \left(\sum_{i=1}^{N}\sum_{j=1}^{N_{\rm IFO}}\frac{\hat{h}_{i,j}}{\sigma_{i,j}^2}\right)\Biggm/\left(\sum_{i=1}^{N}\sum_{j=1}^{N_{\rm IFO}}\sigma^{-2}_{i,j}\right),\label{htot} \\
	\sigma^{\rm tot}=&\left(\sum_{i=1}^{N}\sum_{j=1}^{N_{\rm IFO}}\sigma_{i,j}^{-2}\right)^{-1/2},\label{sigtot}
\end{align}
where $\hat{h}_{i,j}$ is the estimator for the memory amplitude of the $i^{\rm th}$ event detected in the $j^{\rm th}$ interferometer, $\sigma_{i,j}$ is the associated uncertainty, and $\sigma^{\rm tot}$ is the uncertainty associated with $\hat{h}_{\rm tot}$.  The variable, $\sigma_{i,j}$ depends on sky location and polarisation angle.  Combining~(\ref{htot}) and~(\ref{sigtot}) gives an expression for the optimal, total S/N
\begin{align}
	\widehat{{\rm S/N}}_{\rm tot}= \left(\sum_{i=1}^{N}\sum_{j=1}^{N_{\rm IFO}}\frac{\widehat{{\rm S/N}}_{i,j}}{\sigma_{i,j}}\right)\Biggm/\left(\sum_{i=1}^{N}\sum_{j=1}^{N_{\rm IFO}}\sigma^{-2}_{i,j}\right)^{1/2},
\end{align}
where $\widehat{{\rm S/N}}_{i,j}=\hat{h}_{i,j}/\sigma_{i,j}$.

The total expectation value for the total S/N for $N$ events observed with $N_{\rm IFO}$ interferometers is
\begin{align}
	\left<\mbox{S/N}_{\rm tot}\right>=\left(\sum_{i=1}^N\sum_{j=1}^{N_{\rm IFO}}\left<\mbox{S/N}_{i,j}\right>^2\right)^{1/2}.\label{SNtot}
\end{align}
In the limit where the signals from all mergers have the same $\left<\mbox{S/N}\right>_i$, $\left<\mbox{S/N}_{\rm tot}\right>\propto\sqrt{NN_{\rm IFO}}$.  For this analysis we assume a network consisting of LIGO Hanford and Livingston interferometers, $N_{\rm IFO}=2$.

This frequentist approach allows us to estimate the total number of detected merger events required to detect memory.  We also develop a complementary method for determining the Bayesian evidence for a memory signal.  Consider a single coalescence detected by aLIGO.  The Bayesian evidence, $\mathcal{Z}=\int\LL(h|\vec{\xi})p(\vec{\xi})d\vec{\xi}$, where $\vec{\xi}$ are the model parameters, $\LL(h|\vec{\xi})$ is the likelihood of the data $h(t)$ given the model, and $p(\vec{\xi})$ is the prior probability for each of the parameters.   
The log-likelihood is 
\begin{align}
	\ln\LL(h|\vec{\xi})\propto-\frac{1}{2}\sum_{k=1}^M\frac{\left|\tilde{h}_k-\tilde{u}_k/d\right|^2}{\sigma(f_k)^2},\label{LL}
\end{align}
where we sum over frequency bins $f_k$, and $\sigma(f_k)$ is the interferometers noise spectrum.  The variable $\tilde{u}_k(\vec{\xi})$ is the template describing the full merger signal, including both the oscillatory and memory components.  The Bayes factor $\mbox{BF}=\mathcal{Z}/\mathcal{Z}_0$ is the evidence ratio where the denominator is the null hypothesis $\mathcal{Z}_0=\int\LL_0(h|\vec{\xi})p(\vec{\xi})d\xi$.  The likelihood $\LL_0$ is the same as $\LL$ in Eqn.~(\ref{LL}), except we use a different template $\tilde{u}_k^0$ that includes the oscillatory component, but no memory.  The Bayes factor compares the memory hypothesis to the no-memory hypothesis.

For multiple events the evidence is 
\begin{align}
	\mathcal{Z}_{\rm tot}=\prod_{i=1}^{N}\prod_{j=1}^{N_{\rm IFO}}\mathcal{Z}_{i, j}.
\end{align}
The total Bayes factor for $N$ events observed in $N_{\rm IFO}$ interferometers is $\mbox{BF}=\mathcal{Z}_{\rm tot}/\mathcal{Z}_{0,{\rm tot}}$.

Having introduced two statistical formalisms, we apply both to Monte Carlo simulations.  For simplicity, we work with memory-only waveforms, assuming the oscillatory part has been perfectly subtracted. In order to test the validity of this assumption, we derive a `cleaned' memory template by projecting out spectral content covariant with the other astrophysical parameters.  For loud signals, the cleaned template is insensitive to residual errors from imperfect subtraction.  By projecting out part of the memory waveform, we throw out a small amount of signal.  However, we estimate the loss of S/N to be small: only $\sim$0.1\%.

For the sake of pedagogy, we begin with a simulation of an ensemble of binaries with fixed distance, $d=410\,\mbox{Mpc}$ and fixed component masses ($m_1=36\,M_\odot$, $m_2=29\,M_\odot$), but random values of inclination, polarisation, and sky position.  

For each binary we calculate three (expectation values of) signal-to-noise ratios.  Firstly, we calculate the oscillatory signal-to-noise ratio, $\left<\mbox{S/N}_{\rm cbc}\right>$.  Secondly, we calculate the $\psi$-degeneracy-breaking signal-to-noise ratio,  $\left<\mbox{S/N}_{\Delta h}\right>$ (see Eqn.~(\ref{delta_strain})).  We include $\ell\le3$ modes that are key to resolving $\psi$.  Measurement of  $\left<\mbox{S/N}_{\Delta h}\right>>0$ is, in and of itself, interesting as it is evidence of higher-order modes.  We calculate that a single detection of a GW150914-like event at design sensitivity will produce an $\left<\mbox{S/N}_{\Delta h}\right>\gtrsim5$ detection, suggesting this effect can be detected well before design sensitivity \cite[see also][]{capano14,graff15,calderonbustillo16}.  Thirdly, we calculate the memory signal-to-noise ratio $\left<\mbox{S/N}\right>$.  We only retain confident oscillatory detections, $\left<\mbox{S/N}_{\rm cbc}\right>\ge12$~\cite{abadie12c}.  

\begin{figure}
\begin{center}
	\includegraphics[width=0.89\columnwidth]{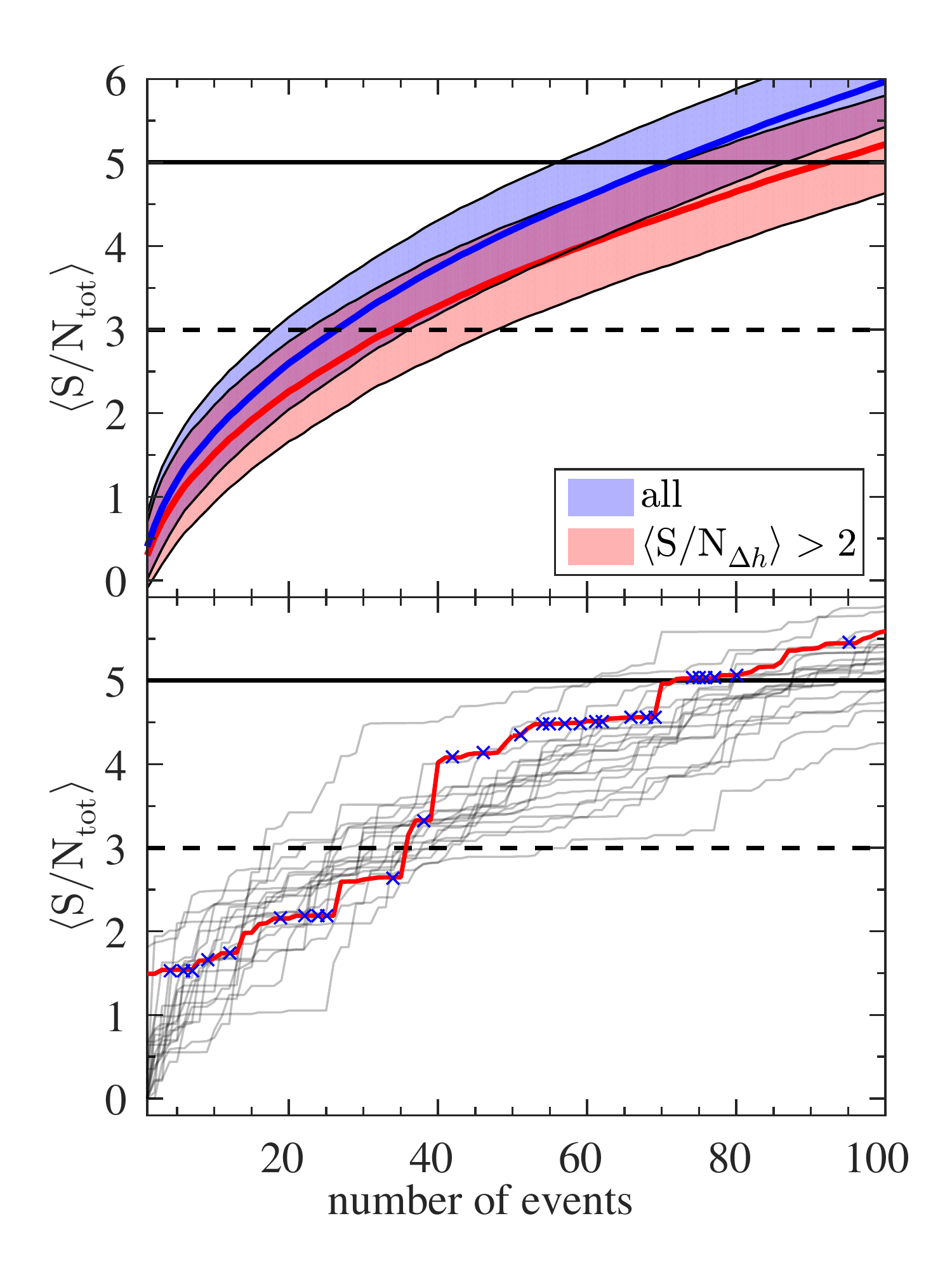}
	\caption{\label{cum_SNR} Evolution of the cumulative signal-to-noise $\left<\mbox{S/N}_{\rm tot}\right>$ as a function of the number of binary black hole mergers.  All binaries have the same distance and mass as the maximum likelihood parameters of GW150914, but have random distributions of inclination, polarisation and sky position.  In the top panel, the solid curves represent the expectation value and the shaded region is the one-sigma uncertainties.  The blue curve sums the memory signal-to-noise contribution from {\it all} binaries, and the red curve assigns memory $\left<\mbox{S/N}\right>=0$ for those binaries where the polarisation angle, and hence the sign of the memory cannot be determined.  The bottom panel shows 20 individual realisations of the red curve in the top panel.  One particular realisation is highlighted in red; the binaries assigned $\left<\mbox{S/N}\right>=0$ are shown with blue crosses.  In both panels, the horizontal dashed and solid lines show $\left<\mbox{S/N}_{\rm tot}\right>=3$ and $5$ respectively.}
\end{center}
\end{figure}

The cumulative $\left<\mbox{S/N}_{\rm tot}\right>$ is shown in Fig.~\ref{cum_SNR}.  In the top panel, the solid curves represent the expectation value while the shaded region is the one-sigma uncertainty.  The blue curve sums the memory contributions from all binaries.  This is unrealistic as it includes binaries where we cannot measure the polarisation, and therefore do not know the memory sign.  The red curve adds the memory contribution only from binaries where we are confident the memory sign is correct.  That is, we only add the memory contribution for signals with $\left<\mbox{S/N}_{\Delta h}\right>>2$, implying we are $\gtrsim95\%$ confident that $\psi$ is accurately measured, and the sign of the memory is correct.  We have verified through simulations\footnote{We perform a Monte Carlo study with GW150914-like binary mergers at a distance such that $\left<\mbox{S/N}_{\Delta h}\right>=2$ with aLIGO sensitivity. We compute the maximum likelihood using templates with the correct sign of the memory, and with the opposite sign, finding that the larger likelihood gives the correct sign of the memory for 95\% of binaries.} that we recover the correct sign of the memory 95\% of the time for signals with $\left<\mbox{S/N}_{\Delta h}\right>=2$.  Binaries that fail this cut are added with memory $\left<\mbox{S/N}\right>=0$.

The bottom panel of Fig.~\ref{cum_SNR} shows 20 Monte Carlo realisations, highlighting the stochasticity of $\left<\mbox{S/N}_{\rm tot}\right>$ growth, and the contribution of the second cut.  We highlight one realisation in red, and show with blue crosses binaries with $\left<\mbox{S/N}_{\Delta h}\right><2$, and therefore have zero memory contribution.

Figure~\ref{cum_SNR} shows that one can expect an $\left<\mbox{S/N}_{\rm tot}\right>=3\,(5)$  detection of memory after $\sim35\,(90)$ GW150914-like detections with aLIGO at design sensitivity, although this could happen with as few as $\sim20\,(75)$.  

\begin{figure}
\begin{center}
	\includegraphics[width=0.9\columnwidth]{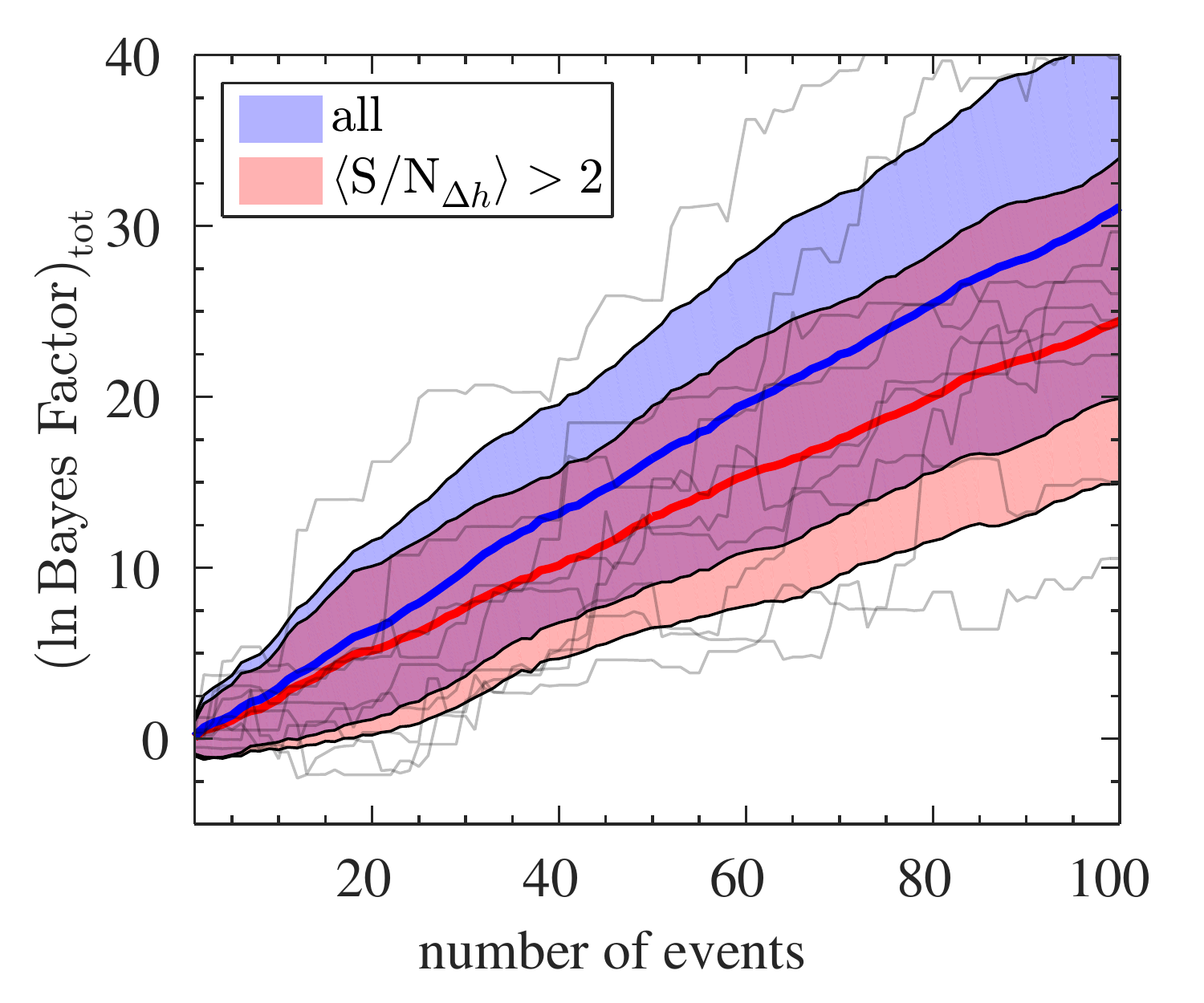}
	\caption{\label{cum_BF} Evolution of the cumulative Bayes factor as a function of the number of binary black hole mergers.  All binaries have the same distance and mass as the maximum likelihood parameters of GW150914, but have random distributions of inclination, polarisation and sky position.  The thick, solid curves represent the expectation value and the shaded region is the one-sigma uncertainties.  The blue curve sums the memory signal-to-noise contribution from {\it all} binaries, and the red curve assigns memory $\left<\mbox{S/N}\right>=0$ for those binaries where the polarisation angle, and hence the sign of the memory cannot be determined.  We also show in grey 10 individual realisations from the red curve.}
\end{center}
\end{figure}

In Fig.~\ref{cum_BF} we plot the cumulative Bayes factor of the memory signal as a function of the number of events.  As with Fig.~\ref{cum_SNR}, we plot in blue the cumulative memory signal from all the binaries, and in red we only add the memory contribution from signals where we are confident that the sign of the memory signal is correct.  Again, the thick curves show the mean value and the shaded region is the one-sigma uncertainties.  As before we highlight the stochastic nature of the growth of this signal by showing 10 individual realisations.  The results of Fig.~\ref{cum_BF} are consistent with that of Fig.~\ref{cum_SNR}: one is likely to be confident of a detection of memory after $\sim35$ events when $\left(\ln\mbox{BF}\right)_{\rm tot}\gtrsim8$

Repeating the simulations presented in Figs.~\ref{cum_SNR} and \ref{cum_BF}, but assuming that events are distributed uniformly in volume, we find that the time to detection changes by less than a few percent.  This is because the growth of $\left<\mbox{S/N}\right>_{\rm tot}$ is dominated by a relatively small number of loud events.  

Given there is only a single GW observation to date, we do not know the mass distribution of binary black holes throughout the Universe.  The memory component of the GW strain scales proportionally to the mass of the binary; if GW150914 was a relatively high-mass binary compared to the population, then the number of events required to detect memory increases.  However, there are some theoretical suggestions \cite[e.g.,][]{demink16} that GW150914 may be at the lower end of the mass distribution, implying GW memory could be detected sooner.  

We provide a proof-of-principle that LIGO and the global network of ground-based GW interferometers will be able to detect GW memory with dozens of nearby events.  The addition of more GW detectors such as Virgo, KAGRA or LIGO-India will further reduce the time to detection.

\begin{acknowledgments}
We are grateful to Lydia Bieri, Ilya Mandel, Chris Matzner, Ed Porter, Vivien Raymond, Letizia Sammut, and Rory Smith for valuable conversations.  We thank the referees for their thorough review of our manuscript.  PDL and YL are supported by an Australian Research Council Discovery Project DP1410102578.  ET and YL are respectively supported through ARC FT150100281 and FT110100384.  YC is supported by NSF Grant No. NSF.1404569.    
\end{acknowledgments}

\bibliography{memory}

\label{lastpage}

\end{document}